# Cryogen-free modular scanning tunneling microscope operating at 4-K in high magnetic field on a compact ultra-high vacuum platform

Angela M. Coe, Guohong Li, and Eva Y. Andrei [a)]

**AFFILIATIONS:**

Department of Physics and Astronomy, Rutgers, the State University of New Jersey, Piscataway, New Jersey 08854, USA

[a)]Author to whom correspondence should be addressed: eandrei@physics.rutgers.edu

(Date: 2 April 2024)

**ABSTRACT:**

One of the daunting challenges in modern low temperature scanning tunneling microscopy (STM) is the difficulty of combining atomic resolution with cryogen free cooling. Further functionality needs, such as ultra-high vacuum (UHV), high magnetic field, and compatibility with µm-sized samples, pose additional challenges to an already ambitious build. We present the design, construction, and performance of a cryogen free, UHV, low temperature, and high magnetic field system for modular STM operation. An internal vibration isolator reduces vibrations in this system allowing atomic resolution STM imaging while maintaining a low base temperature of ~4K and magnetic fields up to 9T. Samples and tips can be conditioned *in-situ* utilizing a heating stage, an ion sputtering gun, an e-beam evaporator, a tip treater, and sample exfoliation. *In-situ* sample and tip exchange and alignment are performed in a connected UHV room temperature stage with optical access. Multisite operation without breaking vacuum is enabled by a unique quick-connect STM head design. A novel low-profile vertical transfer mechanism permits transferring the STM between room temperature and the low temperature cryostat.

## I. INTRODUCTION

The global helium scarcity crisis[1-6] has far reaching consequences in low temperature research due to shortages and prohibitive cost. Liquid helium is indispensable for achieving the low temperatures in scanning tunneling microscopy[7-18]. Low temperature scanning tunneling microscopes (LT-STMs)[19-21] are necessary for atomic-resolution topography of surfaces, for accessing the electronic band structure via scanning tunneling spectroscopy (STS), and for building nanoscale structures.

In STM, a metallic tip is brought within tunneling range (a few nm) of a conductive sample surface and the tunneling current and tip-sample distance are controlled and monitored while the tip is scanned across the sample or held at a fixed location. LT operation improves stability of the tip-sample junction by minimizing thermal drift and enhancing the spatial and energy resolutions by reducing thermal broadening effects. Many material and electronic phenomena can only be explored at LT[22-25]. Most LT-STMs utilize a bath or continuous flow cryostat which relies on a consistent supply of liquid helium to maintain base temperature[26-29]. These systems require frequent refills of liquid helium, forcing experimental interruptions, and subjecting research to the mercy of the helium supply chain.

These problems can be addressed by adapting cryogen-free closed-cycle cryostat (CCC) systems to STM operation. In a CCC, a pulse tube or Gifford-McMahon cryocooler compresses and recycles a small amount of helium gas to achieve LT. No gas refills are needed and no liquid helium is utilized. CCCs provide significant advantages over liquid systems, offering uninterrupted experimental run time, stable LT, and reducing the cost of operation. Unfortunately, CCCs introduce high levels of mechanical vibration (µm level) and acoustic noise generated by the cryocooler, interfering with sensitive STM techniques. Since STM operation requires a very low noise environment (pm levels), low temperature operation in a CCC necessitates a strong thermal link while at the same time being mechanically decoupled from the cryocooler noise. This trade-off is difficult to achieve, evidenced by the scarcity of cryogen-free STM systems available to date[30-39].

Incorporating additional functionalities into the CCC-LT-STM, such as ultra-high vacuum (UHV), high magnetic fields (HF) and µm-sized 2D sample compatibility, provides additional desirable benefits, but comes with added complications. UHV offers a clean environment enabling long term, reliable sample study without contamination; it provides the opportunity to study surface sensitive samples[40,41] that would otherwise degrade in ambient conditions, and is an ideal environment for delivering tip and sample conditioning. However, UHV integration means overcoming added noise from vacuum pumps and requires further separation of the STM from the cooling source, reducing thermal cooling capabilities at LT. Additionally, construction materials must be carefully selected for UHV and wiring techniques have to be UHV specific.

Superconducting magnets allow exploration of vortex phases[7,42], Landau Level spectroscopy[23-25], orbital magnetism[43], and spin polarized quantum states[8]. Incorporation considerations include maintaining the magnet at a sufficiently LT to operate and centering the STM within the magnet. Consideration must be given to the fact that the magnet bore diameter imposes a severe spatial limitation within the LT region.

Another complication of utilizing a superconducting magnet within a system is that optical access is unavailable. This creates a challenge for studying μm-sized samples, including extensively explored 2D materials. To locate a μm-sized sample, the STM tip has to be optically aligned to a gold electrode before being capacitively navigated[44] to the μm-sized sample area. For the system to permit μm-sized sample study, the design must incorporate a way to optically align the tip and sample.

Here we report the development of a CCC-UHV-LT-HF platform which provides variable temperature STM operation from room temperature (RT), down to base temperatures of ~4.6K, in magnetic fields up to 9T. Tip and sample conditioning tools are incorporated into the system to provide high temperature annealing, ion gun sputtering, e-beam evaporation, sample exfoliation, and tip treatment. This instrument uses a compact, modular, quick-connect STM head[45] that is transferable, which solves the optical access problem. This enables tip and sample exchange, alignment, and testing in RT with optical access before being transferred to a LT region with magnetic fields without breaking vacuum. A satellite configuration of the individual chambers together with specialized vacuum transfer methods enable easy movement of the STM head between chambers and precise manipulation of the STM head within a chamber. The system incorporates a unique long range low-profile vertical transfer method[46] to move the STM head to the bottom of a top-loading CCC within a magnet bore. An internal vibration isolation stage[47] and triple wall vacuum design are used within the CCC to maximize cooling while minimizing vibration transfer to the STM. System performance was demonstrated on a highly oriented pyrolytic graphite (HOPG) sample and on a μm-sized graphene sample.

## II. UHV SYSTEM PLATFORM
### A. Overview

The CCC-UHV-LT-HF microscope system consists of five interconnected vacuum chambers[48] arranged in a satellite configuration, in which a large chamber is radially surrounded by four chambers, [Fig. 1]. The center chamber (RTTA) is used for distributing items radially between the surrounding satellite chambers. The satellite chambers are the Load-Lock chamber, the Preparation chamber, the RT-STM chamber, and the Central Hub chamber. The bottom of the Central Hub chamber is connected to a UHV insert that extends into the CCC. This satellite configuration creates a compact design and simplifies transfer between chambers.

The microscope system has a small footprint with dimensions of 1.42m wide, 2.78m long, and 2.24m tall. It is housed on the ground floor of a building within a sound-proofed room to reduce external acoustic vibrations. The controllers for driving the STM and collecting data are located outside the microscope room. Adjacent to the microscope system room is a mechanical room housing gas tanks, a helium compressor[49], a water chiller[50], and a scroll pump[51]. The mechanical room is sound-proofed to contain the acoustic noise of the compressor, chiller, and pump.

The STM head[45] has a quick-connect electrical socket which enables transfer within a vacuum system or between systems, [Fig. 2(a)]. The STM head operates by insertion into an electrical connector plug. The system described here has two connector plugs, one mounted at RT and the other mounted within the CCC at LT. The STM head is compatible with three attachment methods that allow vacuum transfer: puck-style groove, flag-style, and a custom transfer plate. The STM head is compact for ease of transferability and stability with a height of 55.5mm and maximum diameter of 36mm. Full details of this STM head are published elsewhere[45].

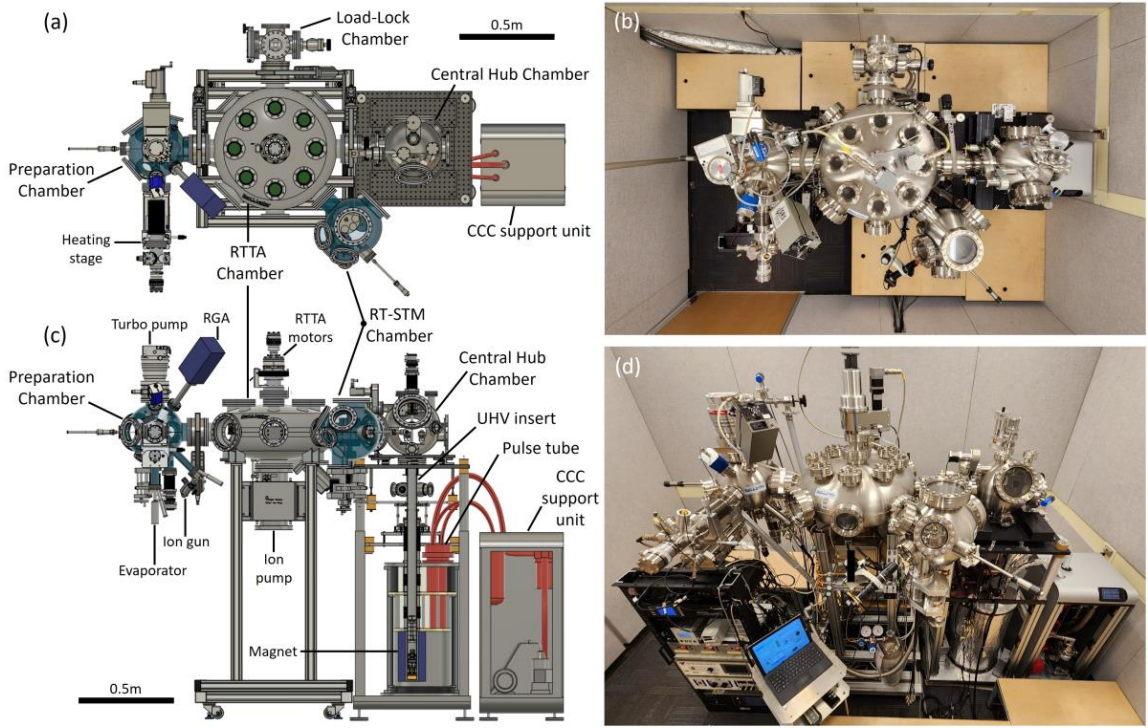

**FIG. 1:** Overview of CCC-UHV-LT-HF microscope system showing the CCC and five chambers: Load-Lock, RTTA, Preparation, RT-SPM, and Central Hub. (a) Top view schematic of system depicting the satellite configuration. (b) Top view photograph of system. The door in the bottom of the photograph leads to the data collection room. The cables and lines running along the top of the photograph exit the microscope room and enter the mechanical room. (c) Front view schematic of system. The UHV insert and CCC are shown as a cross-section. (d) Front view photograph of system.

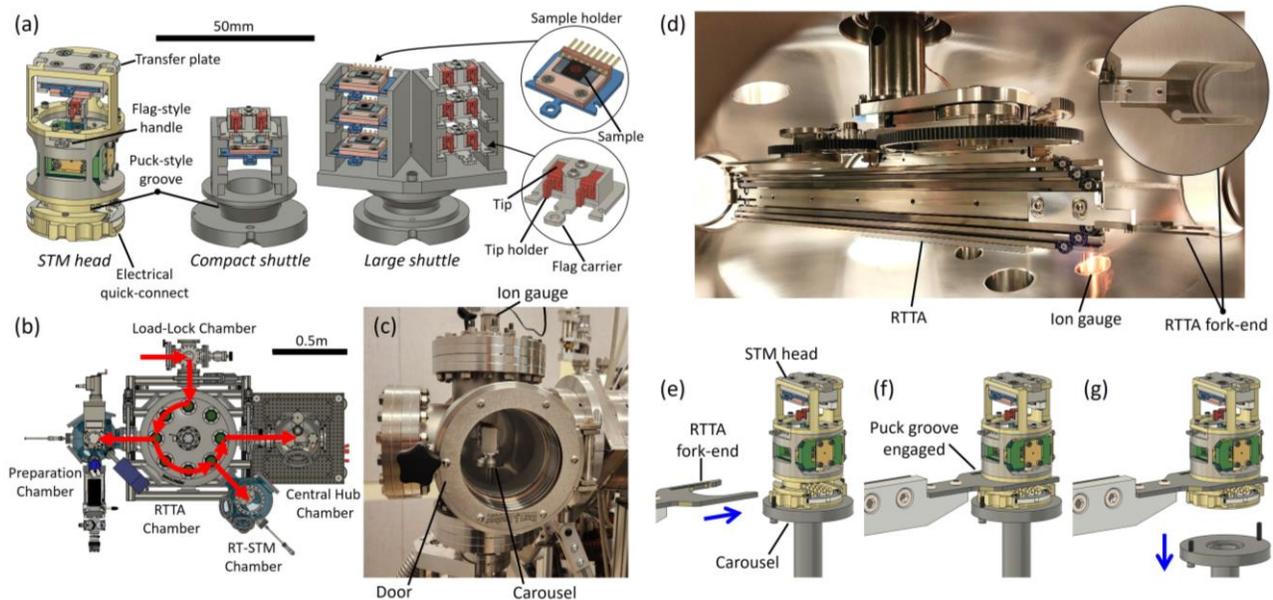

FIG. 2: Transferable items, loading into system, and transfer between chambers. (a) Schematic drawings of items that can be transferred withing the vacuum system. These items include the STM head and transfer shuttles that can hold sample flags and tips in flag carriers. Two versions of the transfer shuttles are shown, a compact and a large size. These shuttles carry the sample and tip flag-holders. (b) Schematic of system with red arrows marking path of travel between chambers using the RTTA. (c) Photograph of Load-Lock chamber. The Load-Lock door and internal carousel are indicated. One of the transfer shuttles is sitting on the carousel. (d) Photograph of RTTA chamber through a window port showing RTTA and fork-end effector for picking up items. The inset photograph shows the fork-end effector. (e-g) Schematics depicting the process of transferring the STM head from the Load-Lock carousel to the RTTA fork-end. (e) The STM head sits on the Load-Lock carousel and the RTTA fork approaches puck-style groove at the base of the STM head. (f) The RTTA fork engages with the STM head groove. (g) The carousel is lowered to disengage with the STM head.

The following provides a detailed walk-through of the platform, demonstrating the functionalities of each chamber as though one were setting up an experiment. Various items, such as samples, tips, STM heads, and auxiliary tools [Fig. 2(a)], are introduced through the Load-Lock chamber to be transferred about the entire system, [Fig. 2(b)]. Intra-chamber transfer is enabled by puck-style grooves in the items. Items that do not have grooves, such as samples and tips mounted to flag-style holders, are loaded onto grooved shuttles for transport between chambers. Each satellite chamber has its own carousel used for transfer and storage of items. The carousel, a circular plate with designated sites to hold items, can rotate[52] (360°) and vertically translate[53] (50.8mm) via a connection to stages controlled from outside the chamber.

### B. Load-Lock Chamber

The Load-Lock chamber has a door for transferring items *in-situ* and has a dedicated pumping line[54] which is used to reach an optimal vacuum level[55] before opening a gate valve[56] separating the Load-Lock from the RTTA chamber [Fig. 2(c)]. Items introduced in the Load-Lock chamber are placed on a carousel for further transfer.

### C. RTTA Chamber

After opening the gate valve, items are transferred out of the Load-Lock chamber and between chambers using a radial telescoping transfer arm (RTTA)[57] equipped with a fork-end effector, [Fig. 2(d)]. The RTTA is mounted to the top of the chamber and can rotate 360° and translate 760mm in-plane to access all chambers. To pick up items, the RTTA is rotated to face the chamber and the fork-end is extended into the chamber [Fig. 2(e)]. The chamber carousel height is adjusted until the RTTA fork is engaged with a puck-style groove in the item [Fig. 2(f)]. The carousel is then lowered to disconnect from the item, [Fig. 2(g)], and the RTTA is retracted into the RTTA chamber and rotated to the desired chamber to deposit the item. The RTTA chamber has an attached ion pump[58] (400L/s), used to pump all chambers to a base pressure of $7 \times 10^{-10}$ torr. The RTTA chamber is structurally supported by slotted framing with no vibration damping. Beams built off this slotted framing provide support to the satellite chambers.

### D. Preparation Chamber

Samples and tips are transferred to a carousel in a dedicated Preparation chamber, [Fig. 3(a)], through the RTTA. A suite of sample and tip conditioning and manipulation equipment is incorporated into this chamber which is accessed without breaking vacuum and isolated to prevent contamination. A wobble stick[59] with a gripper-jaw can move the samples and tips within the preparation chamber by grabbing their flag-style holders. Samples and tips are placed in a heating stage[60] receptacle, which can heat items up to 800C and flash heat to 900C. The stage can rotate and translate to face a low energy ion sputtering gun[61] for cleaning or ion treatments (30eV-1keV), an e-beam evaporator[62] for material deposition, and a custom tip treater for field emission and electron beam bombardment. Bulk crystal samples can be exfoliated *in-situ* within the Preparation chamber and the RT-STM chamber [Fig. 3(b)]. This exfoliation is accomplished by wrapping a flag-style holder with a protruding surface in low-outgassing tape[63,64] and using the wobble stick to touch the tape to the sample surface to remove a few layers. A residual gas analyzer[65] and ion gauge[66] monitor the conditioning process. During conditioning, the preparation chamber is isolated from the other chambers by closing a gate valve[56]. The chamber has a dedicated water-cooled turbo pump[67] (260 L/s), backed by a remote scroll pump[51] located in the mechanical room.

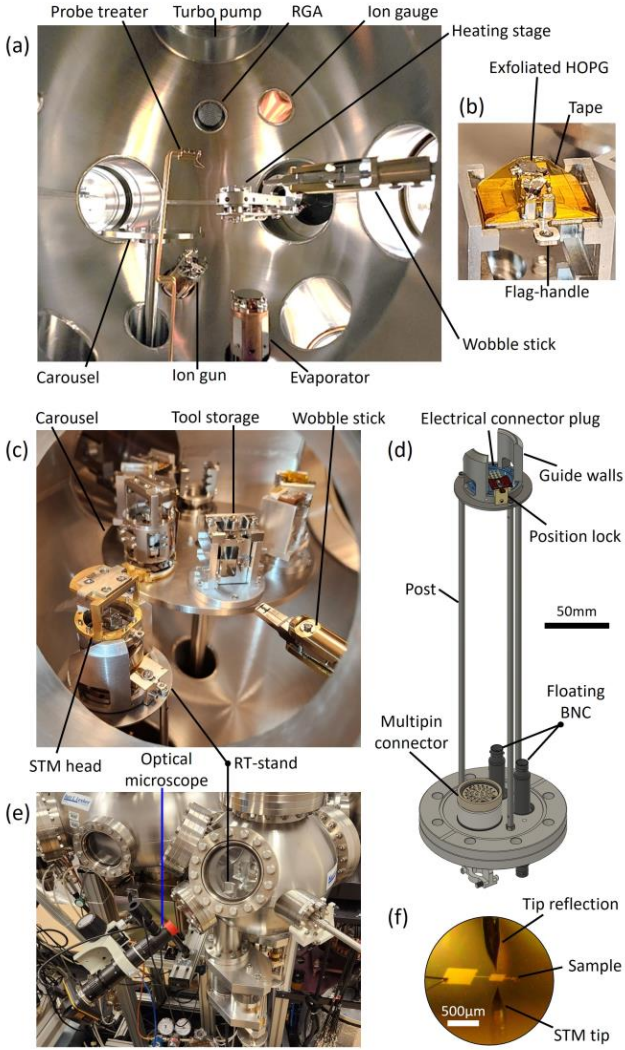

**FIG. 3:** Preparation chamber and RT-STM chamber. (a) Photograph of Preparation chamber through a window port showing the e-beam evaporator, ion sputtering beam, heating stage, tip treater, RGA, ion gauge, turbo pump, carousel, and wobble stick. (b) Photograph of sample exfoliator shown after exfoliating HOPG *in-situ*. (c) Photograph of RT-STM chamber through a window port showing the RT-stand, carousel, tool storage, and wobble stick. The STM head is inserted into the connector plug of the RT-stand. (d) Schematic of RT-stand with the connector plug and electrical vacuum feedthroughs indicated. (e) Photograph of the optical microscope outside the RT-STM chamber focused on the RT-stand. (f) Photograph taken by the optical microscope camera showing the tip and sample alignment on the gold electrode of a μm-sized sample.

### E. RT-STM Chamber

To begin experimentation, the sample, tip, and STM head are transferred to the RT-STM chamber [Fig. 3(c)]. This chamber is used for sample and tip exchange, alignment, and RT testing. A wobble stick moves the STM head from the carousel to a RT-stand. The stand is composed of the STM electrical connector plug mounted to three metal posts (~300mm in length) connected to a vacuum flange with electrical feedthroughs, [Fig. 3(d)]. The vacuum flange is connected to the RT-STM chamber by a translation stage[68] that allows the RT-stand to vertically translate 50.8mm, which is used to provide the wobble stick access to multiple areas of the STM head. The electrical feedthroughs are two floating BNCs[69] for the tunneling and sample bias lines and a multipin connector[70] for the high and low voltage wire lines[71]. The STM head is locked into position on the stand to provide support during sample and tip exchange. The lock is composed of guide walls that surround the STM head to control its lateral movement and a moving component with a flag handle, operated by the wobble stick, to fix its vertical movement. The noise level achieved in this configuration is sufficiently low to permit atomic resolution on HOPG at RT, [Fig. 9(a)]. Once the STM head is locked in place, the wobble stick can perform a tip exchange by grabbing a flag-based tip holder and plugging the tip into the STM head piezoelectric scanner. A sample flag holder can be plugged into the top of the STM head with the wobble stick, completing the STM head assembly process. A tool storage unit is fixed to the RT-STM carousel to hold auxiliary items. An Allen wrench with a flag-style end is stored there and is used for *in-situ* adjustment of the STM head coarse motors via the wobble stick. The tip and sample are aligned using an optical microscope[72] mounted *ex-situ*, [Fig. 3(e)]. This alignment step is critical for studying μm-sized samples before beginning a navigation process[44], [Fig. 3(f)].

### F. Central Hub Chamber

For measurements requiring LT, variable temperature, or high magnetic fields, the RTTA transfers the assembled STM head after tip-sample alignment to the Central Hub chamber, [Fig. 4(a)]. This chamber is the connection between the RT system and the LT region. A gate valve[56] and edge-welded bellows[73] separate the Central Hub chamber from the RTTA chamber. A UHV insert[74] is connected to the bottom of this chamber, which extends into the CCC. The second STM connector plug is attached to the end of the UHV insert on a unique internal vibration isolator[47]. The Central Hub chamber houses a vertical transfer mechanism to move the STM head to the connector plug in the CCC.

### G. Vertical Transfer to CCC

A novel low-profile vertical transfer mechanism[46], housed in the Central Hub chamber, vertically transfers and connects the STM head to the connector plug at the end of the UHV Insert, [Fig. 4(b)]. The mechanism uses a motor-controlled pulley drum mounted in the Central Hub chamber to control the vertical movement of a counterbalance running along two guide-rails[75], [Fig. 4(a)]. The guide-rails extend between the pulley drum to the connector plug at the end of the UHV insert within the cryostat, a distance of 1.3m [Fig. 4(c)]. A carousel in the Central Hub chamber stores two spring-loaded tools, the STM head, and transferable radiation baffles. To begin a vertical transfer, the top of a spring-loaded tool connects to the counterbalance and then the bottom of the spring-loaded tool connects to the top of the STM head, [Fig. 4(a,c)]. One spring-loaded tool is used during the deposition process to transfer the STM head to the bottom of the cryostat, plug it into the connector plug, and disconnect from the STM head to leave it within the cryostat. The other spring-loaded tool is used to recover the STM head after experiments are concluded. The pulley moves the counterbalance, tool, and STM head vertically along the guide-rails between the Central Hub chamber and the CCC. The carousel has clearance for the guide-rails and an open slot to allow the STM head, tool, and counterbalance to pass through it during a transfer. Complete details of the vertical transfer mechanism are outside the scope of this paper and will be published elsewhere.

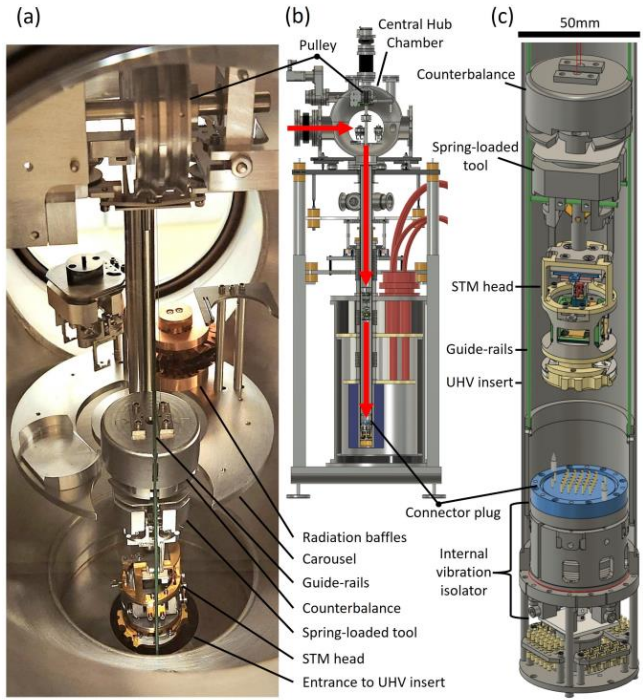

**FIG. 4:** Central Hub chamber and low-profile vertical transfer mechanism. (a) Photograph of Central Hub chamber through a window port showing the vertical transfer mechanism pulley lowering the counterbalance, a spring-loaded tool, and the STM head along the guide-rails into the UHV insert entrance. Carousel is shown storing a second spring-loaded tool and radiation baffles. (b) Schematic cross-section of the Central Hub chamber, UHV insert, and CCC shows the vertical travel path. Red arrow indicates the path the STM head takes to reach the electrical connector plug at the bottom of the UHV insert. Schematic shows the counterbalance, spring-loaded tool, and STM head within the UHV insert enroute to the connector plug. (c) Schematic cross-section of UHV insert bottom end showing electrical connector plug mounted to the internal vibration isolator and the guide-rail connection. STM head is in the process of being lowered to the connector plug.

After the STM head is installed on the connector plug at the end of the UHV insert within the CCC, it must be cooled before low temperature experiments commence.

## III. CLOSED-CYCLE CRYOSTAT
### A. Structure Overview

The primary challenge for STM operation is noise generation by the CCC cryocooler. To provide effective cooling, the cryocooler must be close to the STM head. However, to prevent vibrational noise traveling from the cryocooler to the STM head, they must be mechanically separated. These competing requirements necessitate that the structure connecting the cryocooler and STM head include mechanical isolation while maintaining a strong thermal connection. This is further complicated when including UHV and HF functionality.

In order to solve these challenges, we selected a commercial CCC[76] and modified it for compatibility with UHV and STM. The commercial CCC was originally optimized for operating confocal microscopes and atomic force microscopes in a high vacuum (HV) environment, which are less sensitive to vibrations compared to STMs. This CCC was not compatible with UHV and STM operation. It was selected because of all the commercial options available, this one included a superconducting magnet in the cryostat and reduced the pulse tube's noise. This commercial CCC utilizes a pulse tube cold head[49] with a first stage cooling power of 36W at 40K and a second stage cooling power of 0.9W at 4.2K. The 40K and 4K cooling stages sit within a dewar jacket under HV, and both have a direct connection to the pulse tube [Fig. 5]. The pulse tube operation frequency is 1.4Hz. A support unit, located next to the CCC [Fig. 5(b)], is used for storing the pulse tube's rotary valve and external reservoirs as well as pumps for the dewar. Three flexible hose lines connect the pulse tube to the rotary valve and external reservoirs. The rotary valve connects to the helium compressor, located in the mechanical room. A superconducting magnet[77] is attached to the bottom of the 4K cooling stage with an inner bore diameter of 65.97mm. The magnet bore diameter significantly reduces the working space where the STM head, UHV insert, wiring, vibration damping, and cooling elements must reside. Our modifications to the commercial CCC began with replacing the commercial sample insert with our UHV insert. To obtain a low base temperature, we made modifications to maximize thermal cooling and minimize sources of heating. Also to make the CCC compatible with STM operation, we further reduced the vibration level. The details of these modifications are discussed in the following sections.

To mechanically separate the UHV region from the cryostat, a double-wall-vacuum configuration was created by placing the UHV insert inside a secondary insert[78]. This separates the UHV region from an inner HV region [orange in Fig. 6(b)]. The inner HV volume - a 1mm space between the walls of the two inserts - houses all the wires for the STM head, temperature sensors, and auxiliary lines and their connections to the external controls. This design simplifies the UHV space and enables the use of non-UHV wiring techniques since the only electrical components in UHV are the connector plug and STM head.

The secondary insert sits within a larger outer insert, effectively creating a triple-wall-vacuum configuration. This outer insert is directly connected to the 40K and 4K cooling stages and to the magnet. There is a 15mm space between the secondary and outer inserts, which decreases to 7mm within the magnet. The volume between the secondary and outer inserts is an outer HV volume [grey in Fig 6(b)]. This triple-wall-vacuum arrangement decouples the magnet from the STM, permitting variable temperature STM measurements (4K – 200K) in the presence of high magnetic fields. This allows the magnet to remain below its maximum operating temperature (4.2K), while permitting variation of the STM head temperature.

As mentioned earlier, the connector plug is mounted to the end of the UHV insert within the cryostat [Fig. 5(c)]. The connector plug is mounted on an internal vibration isolator. The UHV insert is sealed with two indium[79] seals, one between the STM connector plug and internal vibration isolator and the other between the isolator and UHV insert. The STM connector plug serves as a vacuum electrical feedthrough for the STM head. Outside the UHV insert, wires run within the HV region between the connector plug at the end of the UHV insert to a feedthrough cross near the top of the UHV insert at RT incorporating a multipin connector vacuum feedthrough and six BNC vacuum feedthroughs. The feedthrough cross is connected to the top of the UHV insert via a flexible bellows, which is used for position alignment. The bottom of the feedthrough cross has a KF connection to the top of the secondary insert. The secondary insert is joined to the top of the outer insert via an ISO connection. The secondary insert and outer insert can be separately pumped to control the two HV regions. These pumps are stored in the CCC's support unit.

Because of the direct connection between the pulse tube and the 40K and 4K cooling stages, the inserts cannot touch the cooling stages and magnet without transferring mechanical vibrations. This makes thermal transfer between the cooling stages and inserts challenging. Further cooling methods were needed to achieve base temperatures of 4.6K at the STM head.

### B. Cooling considerations

To achieve low base temperatures, the cooling power of the cryostat must overcome the thermal load of the STM head, UHV and secondary inserts, wiring, and thermal radiation. The essential aspects that contribute to our low base temperature of 4.6K at the STM head are the use of the triple wall vacuum structure to allow for the use of helium exchange gas outside the UHV region, the STM connector plug cooling, material choice of the inserts and wiring, thermal anchoring at locations of high cooling power, and transferable radiation baffles. Details are given below.

During operation, the STM electrical connector plug is the only thermal conduction pass for cooling the STM head, [Fig. 6(a,b)]. The connector plug is cooled via thermal conduction from the internal vibration isolator and wiring going to the STM. Further cooling of the connector plug is provided by helium exchange gas, which fills the inner and outer HV volumes outside the UHV insert, [Fig. 6(b)]. The bottom of the STM head is made from gold-plated copper, which improves the thermal conduction between the STM head and the connector plug. The connector plug is made from stainless steel for its mechanical strength.

The UHV and secondary inserts are made from thin (~1mm) stainless steel to minimize thermal conduction between RT, the 40K cooling stage, and the 4K cooling stage. To maximize thermal conduction between the cooling stages and inserts, copper fins were welded to the outside of the UHV insert [Fig. 6(c)] and the secondary insert [Fig. 6(d)] near the 40K and 4K cooling stages. These thermal fins are 25mm copper strips (0.05mm thick) that wrap around the diameter of the inserts. The bottom edge of the copper is welded to the inserts while the top edge is fringed and fanned outwards. Vacuum grease[80] was applied to the thermal fins to improve thermal conduction. A higher number of thermal fins are placed at the 40K cooling stage to take advantage of its 36W cooling power. This reduced the thermal load that reaches the 4K cooling stage, which has less cooling power (0.9W). The UHV insert has additional fins at the location of the STM head. The UHV insert fins touch the secondary insert to improve the thermal conduction between them. The secondary insert fins do not touch the outer insert dewar, thus avoiding vibration transfer [Fig. 6(b)]. Where the secondary insert sits within the magnet bore, copper was welded flat to the secondary insert because of space considerations. Despite the secondary fins lack of contact with the outer insert, the thermal fins create a greater surface area with better thermal conduction, improving the cooling efficiency of the helium exchange gas.

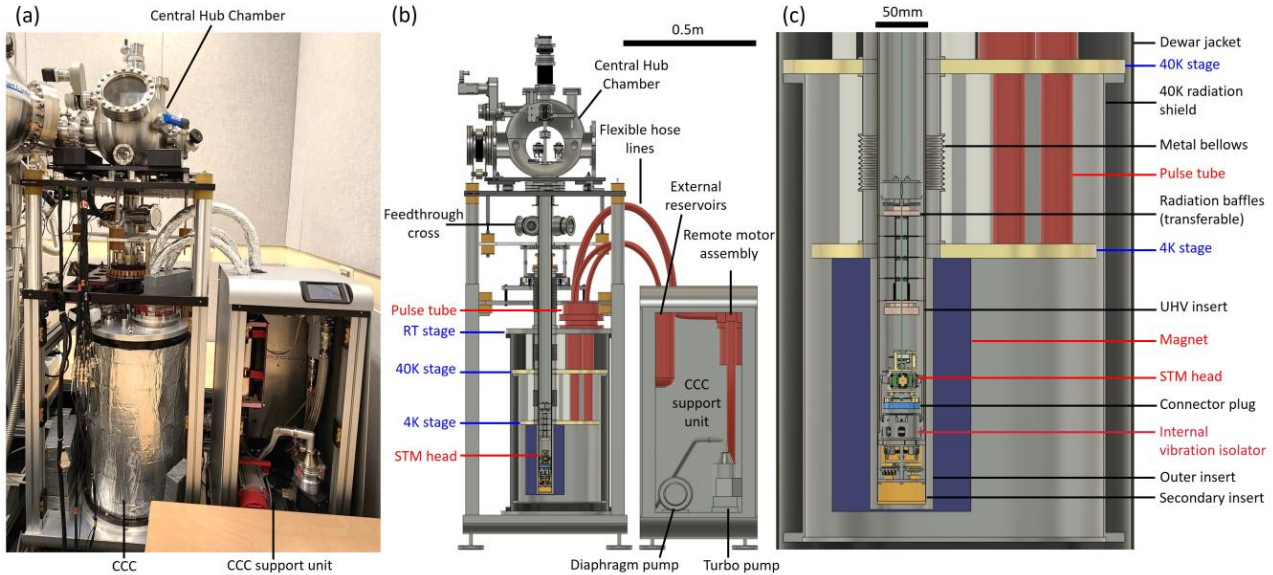

**FIG. 5:** CCC arrangement. (a) Photograph of the Central Hub, UHV insert, and CCC connection. (b) Schematic cross-section showing the Central Hub, UHV insert, and CCC arrangement. (c) Schematic cross-section of region within the CCC cryostat.

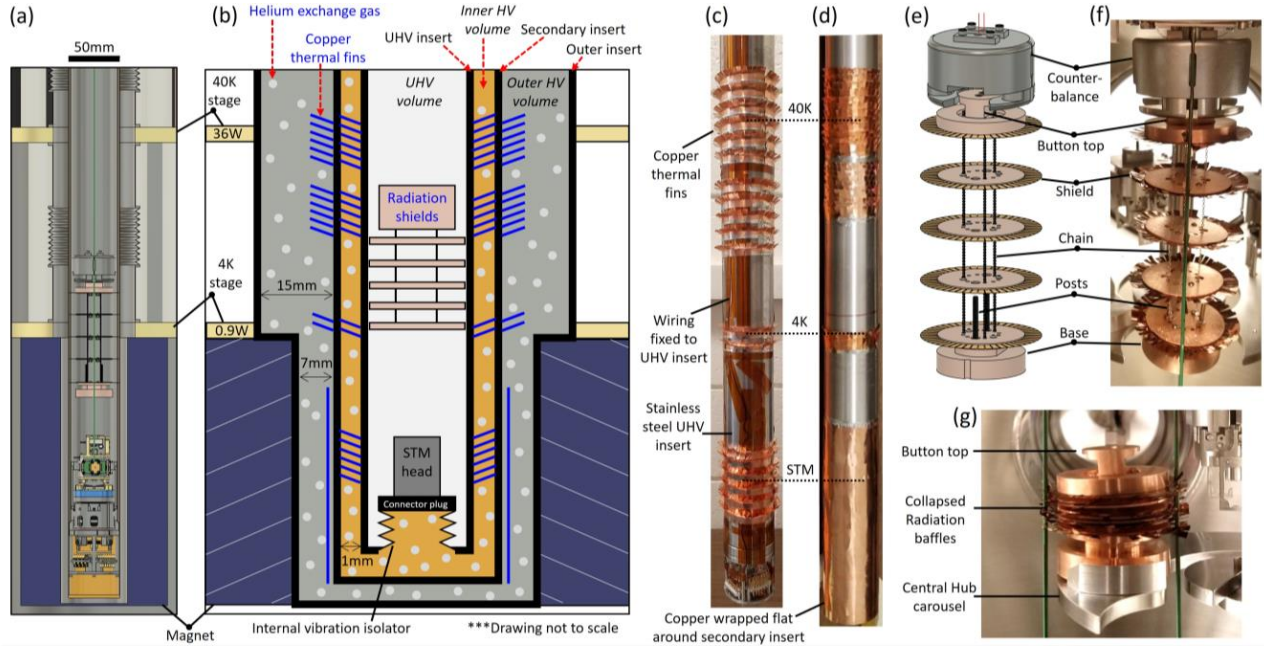

**FIG. 6:** Thermal cooling. (a) Schematic cross-section of STM head within the LT area of the CCC with the 40K cooling stage, 4K cooling stage, and magnet indicated. (b) Cartoon schematic cross-section of LT area of the CCC, not drawn to scale. Illustrates how the UHV insert and secondary insert sit within the outer insert of the CCC with the dimensions between the inserts labeled. The UHV volume (white), inner HV volume (orange), and outer HV volume (grey) are denoted. The copper thermal fins are indicated as blue lines in the cartoon to show their position relative to the cooling stages and magnet. The radiation baffles is positioned within the UHV insert. The STM head is inserted into the connector plug and sits on the internal vibration isolator. Helium exchange gas is used within the inner and outer HV regions (light grey circles). (c) Photograph of the UHV insert showing the wiring path and the copper thermal fins at the 40K cooing stage, the 4K cooling stage, and the STM head. (d) Photograph of the secondary insert showing the copper thermal fins at the 40K and 4K cooling stage positions and copper end wrapping. (e) Schematic and (f) photograph of the radiation baffles held by the vertical transfer counterbalance and ready for transfer into the UHV insert. (g) Photograph of radiation baffles in the collapsed state sitting on the Central Hub chamber carousel before connecting to the pulley counterbalance.

The wiring in the system travels along the outside of the UHV insert, connecting the wire feedthrough cross at RT near the top of the UHV insert to the connector plug located at the bottom of the UHV insert in LT, [Fig. 6(c)]. To prevent the wires from warming the STM head, the wires are thermally anchored at the cooling stages and the wiring material was chosen to minimize thermal conduction while providing an effective electrical signal to operate the STM. Nineteen twisted pair wires are used for the high and low voltage lines of the system. Running between the RT feedthrough cross and 40K stage, the twisted pairs are phosphor bronze[81], selected for its low thermal conductivity. At the 40K cooling stage, the twisted pair material changes to copper for its high thermal conductivity. The copper twisted pairs are thermally anchored at the 40K stage by wrapping them around the UHV insert underneath the copper fins. The twisted pair wires change back to phosphor bronze when leaving the 40K cooling stage to continue to the connector plug at 4K. Six copper coax cables[82] are used for the STM tunneling and sample bias lines, as well as for other secondary lines. These coax cables run along the outside of the UHV insert in a straight path from the RT feedthrough cross to the connector plug at 4K. The coax cables and twisted pairs are fixed to the outside of the UHV insert with Kapton silicone tape[64]. The tape protects the wires from being damaged when the UHV insert is installed in the secondary insert.

Radiation baffles placed within the UHV insert shield the STM head from RT radiative heating coming from the top of the UHV insert. Since the STM head is transferable within the UHV insert via the low-profile vertical transfer mechanism, the radiation baffles were designed with a button top to also be transferable via the counterbalance [Fig. 6(e,f)]. To reduce radiative heat, several requirements had to be met: it was important to have multiple shields to reduce the thermal gradient into smaller steps; the shields had to span a large thermal gradient; each shield had to be thermally anchored to the cooling source but isolated from each other; gaps had to be minimized in the shield to block out radiation from above; and a material of low emissivity (clean, polished) had to be used. To accomplish this, each shield contains two copper discs (36mm diameter) sandwiching a larger (50mm) flexible copper clad Kapton disk which contacts the UHV insert wall (45.8mm) to thermally anchor each shield and minimize gaps between the shields and the UHV insert. To thermally isolate each shield, three flexible stainless-steel chains are used to connect them. To sufficiently span the large thermal gradient, the total height of the radiation baffles was made 116mm when fully extended and hanging from the counterbalance. When not in use, the flexible chains allow the structure to collapse to a height of 41mm for storage on the Central Hub carousel. When collapsed [Fig. 6(g)], the radiation baffles base, shields, and top are held together by two stainless steel posts mounted to the radiation baffles copper base. The radiation baffles assembly is also transferable with the RTTA via a puck-style groove in its base. The temperature of the STM head without the radiation baffles is about 43K and with the radiation baffles is 4.6K. Optimal STM shielding occurs when the vertical transfer mechanism positions the radiation baffles between the 40K and 4K stages. The temperature of the STM can be varied by controlling the position of the radiation baffles within the UHV insert.

**C. Vibration Control**

Tunneling between the tip and sample is exponentially sensitive to their separation. The quality of the STM topography and STS data requires maintaining this tip-sample distance with pm-scale precision. To achieve such conditions, interference from the acoustic, mechanical, and electrical sources between the STM and the environment must be reduced. In a CCC system, the main source of vibrations is from the pulse tube, producing mechanical vibrations at 1.4Hz on the μm-scale. Every structure connecting the pulse tube and STM must be damped to reduce vibrations transferred to the STM head.

The commercial CCC manufacturer has built in elements to reduce some of these vibrations. The CCC has three edge-welded metal bellows [Fig. 5(b,c)] built into the outer insert which help dampen the connection between the 4K cooling stage, the 40K cooling stage, and the secondary insert. Acoustic damping material was used around the support unit. Flexible hose and pumping lines connect between the support unit and cryostat to reduce vibrations. To make this CCC compatible with STM operation and a UHV system, we implemented further vibration reduction methods, [Fig. 7 and 8].

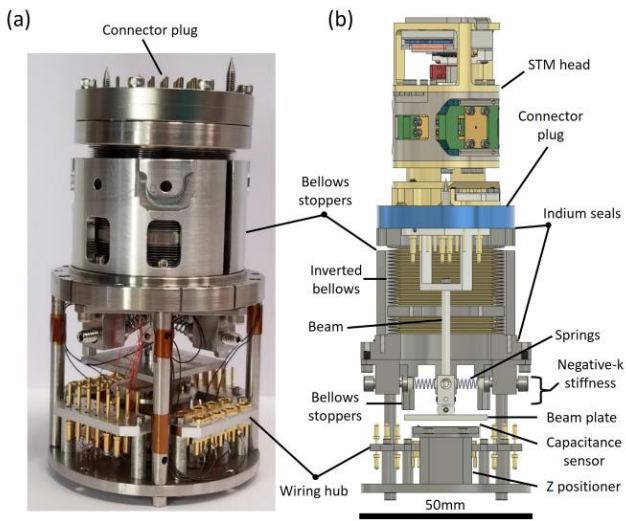

**FIG. 7:** Internal vibration control. (a) Photograph of internal vibration isolator. (b) Schematic cross-section of the internal vibration isolator with STM head inserted into the connector plug.

The primary vibration damping is at the STM head. The STM head through its connector plug is mounted to the top of a novel internal vibration isolator[47] [Fig. 7]. The vibration isolation and damping of this unit are enabled by inverted double stage bellows[83]; negative-k stiffness; and a Z positioner. The electrical connector plug is separated from the UHV insert end via the inverted double stage bellows. The bellows are edge-welded and constructed from 316 stainless steel. This was vital because the bellows had to remain flexible and operate well at low temperatures and stainless-steel was the best choice. The double stage bellows provides two levels of bellows, which act as two band pass filters for the vibration. The shorter lower bellows has a higher resonance frequency, between 20-50Hz. The longer upper bellows has a lower resonance frequency, between 10-20Hz. Overall, the bellows is a low-band pass filter, reducing high frequency vibrations from traveling to the STM head. The rigid design of the STM head permits it to act as a high pass filter, reducing the impact of low frequency vibrations during operation. To prevent over-extension of the bellows during pumping or vertical transfer, two stoppers were installed to limit the bellows range of motion to within a safe range. Outside the UHV insert and within the HV region, the bellows acts with a negative-k stiffness[84,85] structure for further damping. A beam is connected to the connector plug at the top of the bellows and extends down to engage with a negative-k stiffness structure that contains four adjustable compressive springs acting on this beam. The compressive springs act as a counterforce to the bellows to further dampen vibrations. A Z positioner[86] is situated below the bellows in the HV region and can terminate the oscillations by engaging with a plate mounted to the bottom of the bellows beam to dissipate vibrations. A capacitance sensor is located between the bellows beam plate and the top of the Z positioner for monitoring their relative position.

The vast majority of the noise damping is provided by the internal vibration isolator. Additional modifications were made to ensure vibrations coming from other sources were damped before they reach the internal vibration isolator. These are described below.

Polyurethane dampers separate the Central Hub chamber from the dewar structure to reduce vibrations traveling between them. The Central Hub chamber resides on a two-level table, [Fig. 8(a-c)]. The bottom level[87] is fixed to the dewar support structure beams with polyurethane materials separating the beams from the table which proved effective at reducing high frequency vibrations. The top level[88,89] of the table connects to the Central Hub chamber and its position is adjustable, permitting UHV insert and secondary insert alignment within the dewar to prevent internal touching.

To monitor internal touching, the inserts were electrically isolated from the dewar by applying insulating tape to the metal of the ISO connection between the secondary insert [Fig. 8(b)] and outer insert and the pumping line connection to the secondary insert. Internal touching can be monitored via electrical contact resistance while adjusting the top level of the table, which positions the inserts within the dewar. Once the inserts are centered within the dewar, the top and the bottom levels of the table are locked together to create a rigid structure. Additionally, lead weights added to the top table (~104lbs), the base of the dewar support structure (~182lbs), and the top of the pulse tube (~26lbs) were effective at reducing low frequency vibrations, [Fig. 5(a)].

Another vibration modification is eddy current damping[90,91] to reduce vibrations from the feedthrough cross [Fig. 8(b,c)]. The UHV and secondary inserts are suspended in the dewar without internal touching. Testing revealed the feedthrough cross was susceptible to frequencies around 800Hz. To reduce this, the bottom of the UHV insert feedthrough cross is connected to the underside of the two-level table through a damped polyurethane connection, [Fig. 8(b)]. Eddy current damping was implemented below the ISO clamp connection between the secondary insert and outer insert. This consisted of a copper collar attached to the inserts with slots cut into the copper to allow for maximized magnet interaction. The magnets were placed on an adjustable support stage that does not touch the inserts. The adjustable support stage has polyurethane separating it from the dewar support structure.

The turbo pump and its water cooling on the Preparation chamber produce mechanical noise vibrations at ~1kHz and ~100Hz respectively. To prevent these vibrations reaching the Central Hub chamber via the RTTA, a metal bellows with urethane support [Fig. 8(d)] was installed between them. The urethane support prevents the bellows from collapsing when the chambers are under vacuum.

When a large quantity of electronic equipment is being used without a dedicated ground, it is necessary to prevent electrical signals of 60Hz and its harmonics (ground loop). To accomplish this, the ground of the STM control electronics[92] was floated by connecting it to a transformer[93]. The tunneling and sample bias lines on the RT feedthrough cross were floated from the system by applying insulating tape along the metal of the KF connection [Fig. 8(b)]. One circuit breaker with a shared ground powers all equipment in the microscope room. An acoustic damping material was wrapped around the CCC jacket dewar and around the flexible lines connecting the pulse tube to the CCC support unit.

## IV. OPERATION AND PERFORMANCE
### A. Overview

This section demonstrates the performance and operation of the instrument at RT and LT on an HOPG sample and on a μm-sized sample. Data was obtained using a mechanically cut PtIr tip and an RHK R9plus controller.

### B. RT performance

The STM performance in the RT-STM chamber is demonstrated using an HOPG sample. The sample was exfoliated *in-situ* in the Preparation chamber, [Fig. 3(b)]. The tip and sample were installed into the STM head on the RT-stand. The STM head is sufficiently stable to achieve atomic resolution without any vibration isolation on the RT-stand, [Fig. 9(a)].

### C. LT cooldown operation

Once RT measurements are concluded, the RTTA moves the STM head to the Central Hub chamber for transfer to LT. During the vertical transfer process, both the inner and outer HV volumes are filled with exchange gas (>100mbar). This cools the inserts and connector plug while also expanding the internal vibration isolator bellows causing it to become rigid, providing a stable platform for transferring and connecting the STM head. The vertical transfer process takes about 15min, where most of the time is spent traveling between the RT and LT regions. The vertical transfer mechanism is then used to transfer the radiation baffles to a position between the 40K and 4K stages. Cooling the STM head from RT to 4.6K takes about 18hrs. The temperature is monitored from three temperature sensors mounted within the STM head, the bottom of the UHV insert, and the magnet. The cooldown time is shown in [Fig. 9(b)]. The base temperature for the STM head is 4.6K, the UHV insert is 4.2K, and the magnet is 3.9K.

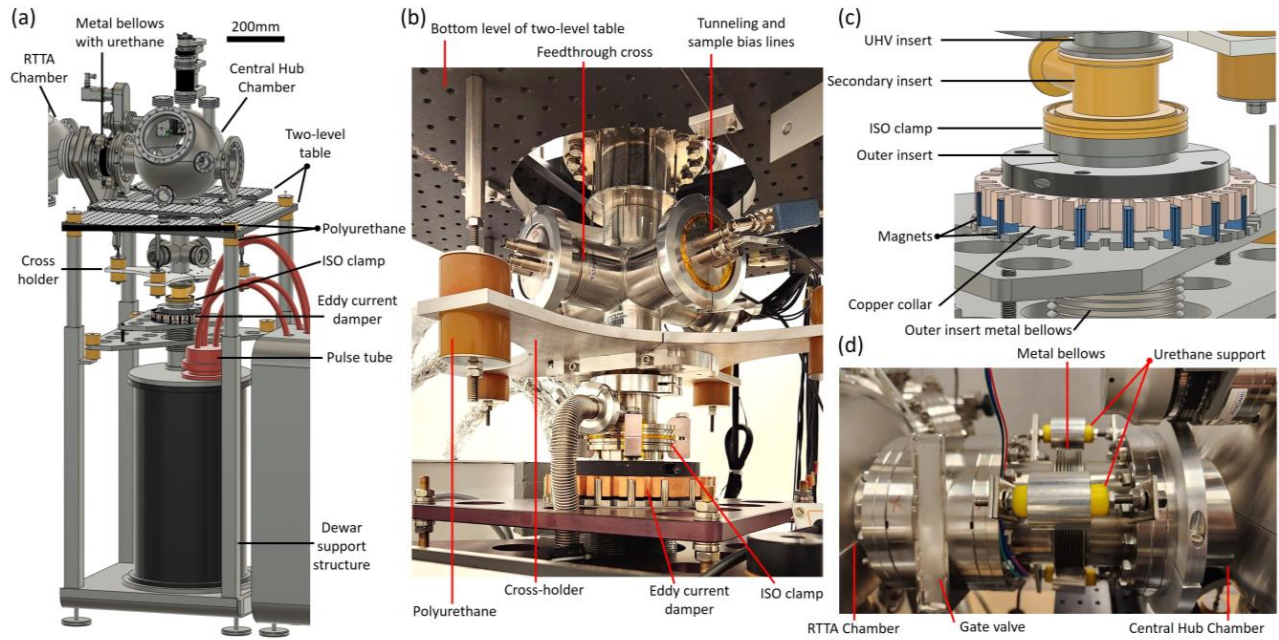

**FIG. 8:** External vibration control. (a) Schematic of CCC assembly with vibration damping elements indicated. (b) Photograph of feedthrough cross and ISO clamp joint between secondary insert and outer insert. Eddy current damper is located beneath the ISO clamp. Tunneling and sample bias feedthrough connection to the cross is shown, where tape (orange) is used to electrically isolate the flange from system. (c) Schematic of eddy current damper and connections between UHV, secondary, and outer inserts. (d) Metal bellows between RTTA and Central Hub chambers with three urethane support structures.

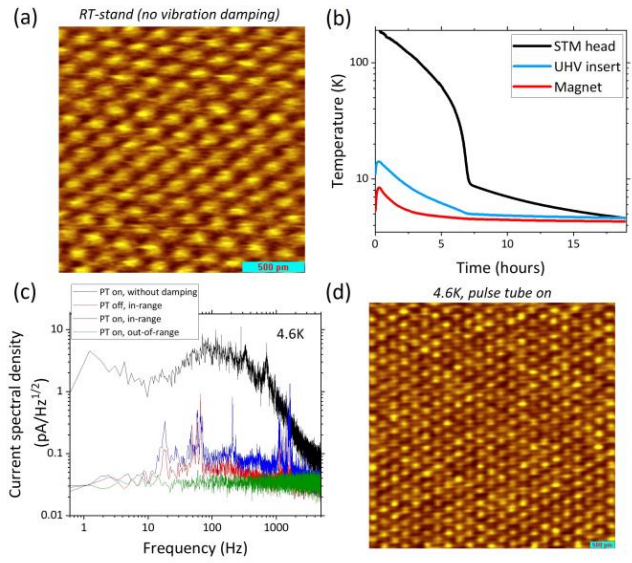

**FIG. 9:** System performance. (a) RT-stand STM topography image of HOPG showing atomic resolution, ($V_{bias}$=200pA, I=200pA). (b) Cool down of STM head after insertion into the connector plug within the CCC. The magnet and UHV insert temperature rise slightly when STM head is first inserted. The temperature of the STM head, UHV insert, and magnet cool to base temperatures of 4.6K, 4.2K, and 3.9K. (c) Current noise in tunneling line with and without vibration isolation. The green curve shows the noise level out of range with the pulse tube (PT) running and vibration isolation active. The blue curve shows the noise level in-range of the sample with pulse tube on and vibration isolation active. The red curve shows in-range tunneling with vibration isolation and the pulse tube off. The black curve shows the noise level in-range with no vibration isolation and the pulse tube on. (d) Atomic resolution achieved on HOPG with pulse tube on at base temperature, ($V_{bias}$=50pA, I=100pA).

Once base temperature is reached and before performing measurements, the helium pressure in the inner volume is reduced to $4.3 \times 10^{-3}$ mbar to make the internal vibration isolator bellows flexible and eliminate the chance of sparking. In this system, experiments can run at stable low temperatures indefinitely. Once the experiment is concluded, the inner volume is refilled with helium exchange gas to stiffen the internal vibration isolator bellows and allow for the vertical transfer mechanism to retrieve the STM.

### D. Base temperature within CCC

The performance of the STM head at base temperature with the pulse tube running is shown in [Fig. 9(c,d)]. The noise in the tunneling line is very sensitive to vibration and can be effectively used to evaluate the vibration noise level at the location of the STM head. The LT tunneling current for the tip within tunneling range (in-range) and out-of-range of the sample are shown in [Fig. 9(c)]. To demonstrate the effectiveness of the vibration isolation, the tunneling current for the tip with and without vibration isolation is also shown in [Fig. 9(c)]. This data was acquired with an open-feedback tunneling setpoint of $V_{bias}$ = 200mV and I =200pA, where $V_{bias}$ is the voltage applied to the sample and I is the dc current measured from the tip, and a cut off frequency of 250Hz was applied. When the vibration isolation is bypassed and the pulse tube is on, the tunneling current shows a large noise level around 10pA/sqrt(Hz). When vibration isolation is implemented, this noise decreases substantially as seen in the other three curves. The out-of-range data is featureless with a background of about 40fA/sqrt(Hz), attributed to the cabling used. When in-range, the current shows no vibration at the pulse tube frequency of 1.4Hz. The overall noise level is below 1.2pA/sqrt(Hz). Only a slight increase is observed when the pulse tube is turned on. Atomic resolution imaging on HOPG was achieved at 4.6K with the pulse tube running as shown in [Fig. 9(d)]. The thermal drift measured at 4.6K over the course of a few days was less than 1nm/day.

### E. 2D μm-sized samples within CCC

This system can be used to study μm-sized samples, such as twisted bilayer graphene (TBG) on hBN. We studied a TBG sample near the magic angle (~1.1° twist angle) that included vacancies created by *in-situ* He ion sputtering [Fig. 10(a)] using the ion sputtering gun in the Preparation chamber. After ion sputtering the sample was plugged into the STM head on the RT-stand and the STM tip was optically aligned to the large gold electrode connected to the sample, [Fig. 3(f)]. The STM head was then transferred to the CCC for LT measurements where the tip was navigated to the sample using a capacitive based method developed by our group[44] without optical access. A topography image of the sample shows a Moiré pattern typical of magic angle TBG, [Fig. 10(b)].

### F. Variable temperature and Magnetic field at high temperature

The temperature of the STM head can be varied between RT and base temperature using multiple methods. These methods include utilizing a local heater within the sample head for heating the sample, moving the radiation baffles within the UHV insert, and changing the inner and outer HV region pressures. The transferable radiation baffles can be moved within the UHV insert which changes the temperature of the STM head between 4.6K to about 43K [Fig. 11(a)]. This is a gentle process that allows the tip to maintain its position over a specific sample area with minimal thermal drift. While varying the STM head temperature, the magnet temperature remains below 4.2K (operational limit), allowing for magnetic fields to be applied at elevated temperatures. This is demonstrated in [Fig. 11(b)] by applying a magnetic field of 5T to a HOPG sample while the STM head is at a temperature of 20K and measuring the dI/dV. The current setpoint is varied between 100pA and 400pA while at 5T.

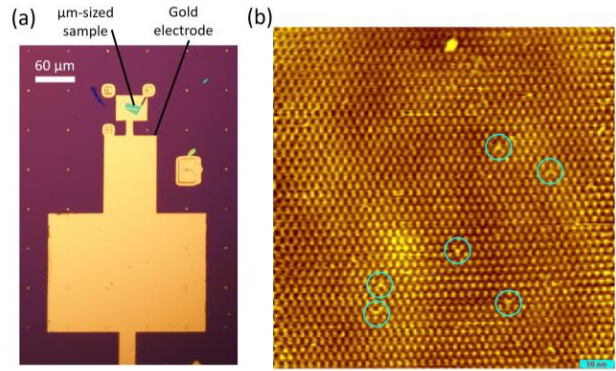

**FIG. 10:** STM head operation with a μm-sized sample in the CCC at base temperature with no optical access. (a) μm-sized sample device with sample area and gold electrode indicated. The sample used was TBG on hBN and sputtered *in-situ* with helium ions. (b) Topography of Moiré pattern with image size 500nm x 500nm, ($V_{bias}$=200pA, I=200pA). Circled features are some of the defects created by helium ion sputtering.

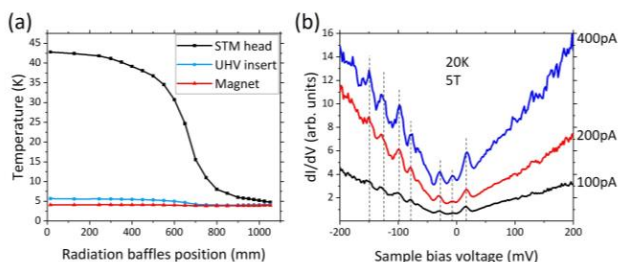

**FIG. 11:** Variable STM temperature with radiation baffles and magnetic field applied at 20K. (a) Radiation baffles position vs temperature of STM head, UHV insert, and magnet. Inner HV region has a helium exchange gas pressure of 4.3E-3mbar and the outer HV region has a pressure of 310mbar. The magnet temperature stays below 4.2K, its operation temperature limit, during temperature variation. (b) Magnetic field of 5T applied to HOPG at 20K showing Landau levels. Current setpoint is varied, (Vbias=200pA, 400kHz sinusoidal modulation of Vrms=5mV).

## V. SUMMARY AND OUTLOOK

We report the successful incorporation of CCC cryocooling with highly sensitive STM operation and UHV-LT-HF. Base temperatures down of 4.6K were achieved and atomic resolution was successfully obtained. The low base temperature was achieved by using a triple-wall vacuum configuration which streamlined the UHV region and enabled the use of strong thermal links. Reduction of noise from the CCC pulse tube was achieved using an internal vibration isolator. Through the use of a quick-connect STM head, optical access is available for RT tip and sample exchange, alignment and measurement before being transferred to the CCC for measurements at LT, variable temperature, and high magnetic fields up to 9T. Tip and sample conditioning is incorporated into the system for *in-situ* use. The satellite configuration of this platform simplifies the transfer between chambers and allows the system to be contained in a small footprint. The compact size of the system is also attributed to the use of a low-profile vertical transfer mechanism used to transfer the STM head from RT to the CCC.

The system is optimized for studying samples ranging in size from mm to μm. Importantly, it allows *in-situ* study of air sensitive samples. With the modular design of the quick-connect STM head, it will be possible to extend the suite of exchangeable probes to include atomic force microscopes (AFM), magnetic force microscopy (MFM), and electrical transport measurements.

## ACKNOWLEDGMENTS

The authors thank the following for helpful discussions and/or technical assistance: Paul Pickard, William Schneider, Eric Paduch, and Cole Woloszyn from the Rutgers Physics and Astronomy Machine Shop. The instrumentation and infrastructure were primarily supported by DOE-FG02-99ER45742, NSF-MRI 1337871, and Gordon and Betty Moore Foundation GBMF9453.

## AUTHOR DECLARATIONS:

A.M.C., G.L., and E.Y.A have Patent No. US 11,474,127 B2 issued. G.L. and E.Y.A have Patent No. WO 2019/209592 A2 pending. A.M.C., G.L., and E.Y.A have Patent No. WO 2021/226354 A1 pending.

## DATA AVAILABILITY:

The data that support the findings of this study are available from the corresponding author upon reasonable request.

## SUPPORTING INFORMATION (SI): NA